\documentclass{webofc}
\usepackage[varg]{txfonts}   
\usepackage{graphicx}   
\usepackage{float}
\usepackage[export]{adjustbox}
\usepackage{caption}
\usepackage{subcaption}
\usepackage{xcolor}
\usepackage{verbatim}   
\usepackage{color}      
\usepackage{hyperref}   
\usepackage{bm} 
\usepackage{multirow}
\usepackage{csquotes}
\begin{document}
	
%

\newcommand{\pp}           {pp\xspace}
\newcommand{\ppbar}        {\mbox{$\mathrm {p\overline{p}}$}\xspace}
\newcommand{\XeXe}         {\mbox{Xe--Xe}\xspace}
\newcommand{\PbPb}         {\mbox{Pb--Pb}\xspace}
\newcommand{\pA}           {\mbox{pA}\xspace}
\newcommand{\AAa}          {\mbox{AA}\xspace}
\newcommand{\pPb}          {\mbox{p--Pb}\xspace}
\newcommand{\AuAu}         {\mbox{Au--Au}\xspace}
\newcommand{\dAu}          {\mbox{d--Au}\xspace}
\newcommand{\CuCu}         {\mbox{Cu--Cu}\xspace}

\newcommand{\s}            {\ensuremath{\sqrt{s}}\xspace}
\newcommand{\snn}          {\ensuremath{\sqrt{s_{\mathrm{NN}}}}\xspace}
\newcommand{\pt}           {\ensuremath{p_{\rm T}}\xspace}
\newcommand{\meanpt}       {$\langle p_{\mathrm{T}}\rangle$\xspace}
\newcommand{\ycms}         {\ensuremath{y_{\rm CMS}}\xspace}
\newcommand{\ylab}         {\ensuremath{y_{\rm lab}}\xspace}
\newcommand{\etarange}[1]  {\mbox{$\left | \eta \right |~<~#1$}}
\newcommand{\yrange}[1]    {\mbox{$\left | y \right |~<$~0.5}}
\newcommand{\yran}[2]      { 0~<~ $ y_{cm} $ ~<~0.5}
\newcommand{\dndy}         {\ensuremath{\mathrm{d}N_\mathrm{ch}/\mathrm{d}y}\xspace}
\newcommand{\dndeta}       {\ensuremath{\mathrm{d}N_\mathrm{ch}/\mathrm{d}\eta}\xspace}
\newcommand{\avdndeta}     {\ensuremath{\langle\dndeta\rangle}\xspace}
\newcommand{\dNdy}         {\ensuremath{\mathrm{d}N_\mathrm{ch}/\mathrm{d}y}\xspace}
\newcommand{\dNdyy}        {\ensuremath{\mathrm{d}N/\mathrm{d}y}\xspace}
\newcommand{\Npart}        {\ensuremath{N_\mathrm{part}}\xspace}
\newcommand{\Ncoll}        {\ensuremath{N_\mathrm{coll}}\xspace}
\newcommand{\dEdx}         {\ensuremath{\textrm{d}E/\textrm{d}x}\xspace}
\newcommand{\RpPb}         {\ensuremath{R_{\rm pPb}}\xspace}
\newcommand{\RAA}         {\ensuremath{R_{\rm AA}}\xspace}
\newcommand{\res}         {\ensuremath{\sigma}\xspace}

\newcommand{\nineH}        {$\sqrt{s}~=~0.9$~Te\kern-.1emV\xspace}
\newcommand{\seven}        {$\sqrt{s}~=~7$~Te\kern-.1emV\xspace}
\newcommand{\twoH}         {$\sqrt{s}~=~0.2$~Te\kern-.1emV\xspace}
\newcommand{\twosevensix}  {$\sqrt{s}~=~2.76$~Te\kern-.1emV\xspace}
\newcommand{\five}         {$\sqrt{s}~=~5.02$~Te\kern-.1emV\xspace}
\newcommand{\twosevensixnn}{$\sqrt{s_{\mathrm{NN}}}~=~2.76$~Te\kern-.1emV\xspace}
\newcommand{\fivenn}       {$\sqrt{s_{\mathrm{NN}}}~=~5.02$~Te\kern-.1emV\xspace}
\newcommand{\LT}           {L{\'e}vy-Tsallis\xspace}
\newcommand{\GeVc}         {Ge\kern-.1emV/$c$\xspace}
\newcommand{\MeVc}         {Me\kern-.1emV/$c$\xspace}
\newcommand{\TeV}          {Te\kern-.1emV\xspace}
\newcommand{\GeV}          {Ge\kern-.1emV\xspace}
\newcommand{\MeV}          {Me\kern-.1emV\xspace}
\newcommand{\GeVmass}      {Ge\kern-.2emV/$c^2$\xspace}
\newcommand{\MeVmass}      {Me\kern-.2emV/$c^2$\xspace}
\newcommand{\lumi}         {\ensuremath{\mathcal{L}}\xspace}

\newcommand{\ITS}          {\rm{ITS}\xspace}
\newcommand{\TOF}          {\rm{TOF}\xspace}
\newcommand{\ZDC}          {\rm{ZDC}\xspace}
\newcommand{\ZDCs}         {\rm{ZDCs}\xspace}
\newcommand{\ZNA}          {\rm{ZNA}\xspace}
\newcommand{\ZNC}          {\rm{ZNC}\xspace}
\newcommand{\SPD}          {\rm{SPD}\xspace}
\newcommand{\SDD}          {\rm{SDD}\xspace}
\newcommand{\SSD}          {\rm{SSD}\xspace}
\newcommand{\TPC}          {\rm{TPC}\xspace}
\newcommand{\TRD}          {\rm{TRD}\xspace}
\newcommand{\VZERO}        {\rm{V0}\xspace}
\newcommand{\VZEROA}       {\rm{V0A}\xspace}
\newcommand{\VZEROC}       {\rm{V0C}\xspace}
\newcommand{\Vdecay} 	   {\ensuremath{V^{0}}\xspace}

\newcommand{\ee}           {\ensuremath{e^{+}e^{-}}} 
\newcommand{\pip}          {\ensuremath{\pi^{+}}\xspace}
\newcommand{\pim}          {\ensuremath{\pi^{-}}\xspace}
\newcommand{\pipm}         {\ensuremath{\pi^{\pm}}\xspace}
\newcommand{\kap}          {\ensuremath{\rm{K}^{+}}\xspace}
\newcommand{\kam}          {\ensuremath{\rm{K}^{-}}\xspace}
\newcommand{\pbar}         {\ensuremath{\rm\overline{p}}\xspace}
\newcommand{\kzero}        {\ensuremath{{\rm K}^{0}_{\rm{S}}}\xspace}
\newcommand{\lmb}          {\ensuremath{\Lambda}\xspace}
\newcommand{\almb}         {\ensuremath{\overline{\Lambda}}\xspace}
\newcommand{\Om}           {\ensuremath{\Omega^-}\xspace}
\newcommand{\Mo}           {\ensuremath{\overline{\Omega}^+}\xspace}
\newcommand{\X}            {\ensuremath{\Xi^-}\xspace}
\newcommand{\Ix}           {\ensuremath{\overline{\Xi}^+}\xspace}
\newcommand{\Xis}          {\ensuremath{\Xi^{\pm}}\xspace}
\newcommand{\Oms}          {\ensuremath{\Omega^{\pm}}\xspace}
\newcommand{\degree}       {\ensuremath{^{\rm o}}\xspace}
\newcommand{\rh}           {\ensuremath{\rm {\rho}^{\rm 0}}\xspace}
\newcommand{\rhmass}       {\ensuremath{\rm {\rho(770)}^{\rm 0}}\xspace}
\newcommand{\kstarmass}    {\ensuremath{\rm {K(892)}^{\rm{* 0}}}\xspace}
\newcommand{\kstarpmmass}  {\ensuremath{\rm {K(892)}^{\rm{*\pm}}}\xspace}
\newcommand{\kstar}        {\ensuremath{\rm {K}^{\rm{* 0}}}\xspace}
\newcommand{\kstarpm}      {\ensuremath{\rm {K}^{\rm{*\pm}}}\xspace}
\newcommand{\sigmass}      {\ensuremath{\rm {\Sigma(1385)}^{\rm{\pm}}}\xspace}
\newcommand{\sigm}         {\ensuremath{\rm {\Sigma}^{\rm{\pm}}}\xspace}
\newcommand{\ximass}       {\ensuremath{\rm {\Xi(1530)}^{\rm{0}}}\xspace}
\newcommand{\xim}          {\ensuremath{\rm {\Xi}^{\rm{0}}}\xspace}
\newcommand{\ximinus}      {\ensuremath{\rm {\Xi}^{\rm{-}}}\xspace}
\newcommand{\lambmass}     {\ensuremath{\rm {\Lambda(1520)}}\xspace}
\newcommand{\lstar}        {\ensuremath{\rm {\Lambda}^{\rm{* }}}\xspace}
\newcommand{\phim}         {\ensuremath{\phi}\xspace}
\newcommand{\phimass}      {\ensuremath{\phi(1020)}\xspace}
\newcommand{\pik}          {\ensuremath{\pi\rm{K}}\xspace}
\newcommand{\kk}           {\ensuremath{\rm{K}\rm{K}}\xspace}
\newcommand{\zero}         {\ensuremath{^{\rm 0}}\xspace}
\newcommand{\kskm}{$\mathrm{K^{*0}/K^{-}}$}
\newcommand{\phikm}{$\mathrm{\phi/K^{-}}$}
\newcommand{\phixi}{$\mathrm{\phi/\Xi}$}
\newcommand{\phiom}{$\mathrm{\phi/\Omega}$}
\newcommand{\xiphi}{$\mathrm{\Xi/\phi}$}
\newcommand{\omphi}{$\mathrm{\Omega/\phi}$}
\newcommand{\kstf} {K$^{*}(892)^{0}~$}
\newcommand{\phf} {$\mathrm{\phi(1020)}~$}
\newcommand{\dd}{\ensuremath{\mathrm{d}}}
\newcommand{\mT}{\ensuremath{m_{\mathrm{T}}}\xspace}
\newcommand{\krr}{\ensuremath{\kern-0.09em}}
\newcommand{\npart}{\ensuremath{\langle N_{\mathrm{part}}\rangle}\xspace}
\newcommand{\ncoll}{\ensuremath{\langle N_{\mathrm{coll}}\rangle}\xspace}
\title{System size and energy dependence of resonance \\production in ALICE}
%
%
\author{\firstname{Vikash} \lastname{Sumberia}\inst{1} (\bf{for the ALICE collaboration})\thanks{\email{vikash.sumberia@cern.ch}}}


\institute{University of Jammu, Jammu and Kashmir, 180006, India }

\abstract{
  Hadronic resonances, thanks to their relatively short lifetimes, can be used to probe the properties of the hadronic phase in ultrarelativistic heavy-ion collisions. In particular their lifetimes are exploited to investigate the interplay between particle rescattering and regeneration after hadronization. In this contribution we present recent results on $\rho(770)^0$, $K^{*}$(892), $\phi$(1020), $\Sigma$(1385)$^{\pm}$, $\Xi(1530)^{0}$ and $\Lambda(1520)$ production in pp, p$-$Pb, \PbPb and \XeXe collisions at LHC energies.
}

\maketitle

\section{Introduction}
\label{intro}
Production properties of hadronic resonances and their comparison with those of stable hadrons are useful tools to understand the properties of the strongly interacting matter formed in relativistic nucleus-nucleus collisions. Due to their short lifetimes comparable to the lifetime of the hadronic phase~\cite{1}, resonances provide an insight into the dynamical evolution of the fireball. The estimated yield of the resonances may be modified even after the chemical freezeout due to (pseudo)-elastic rescattering or regeneration of their decay daughters and these processes are dominant at low \pt ($\lesssim$ 2 $\xspace$ \GeVc)~\cite{2,3,4}. The rescattering process supresses the observed yields with respect to what is produced at chemical freeze-out while regeneration enhances the yield of resonances~\cite{5}. Thus, the study of resonance particles can provide a hint for finite lifetime of the hadronic phase (resonance to stable hadron yield ratios), understand the partonic in-medium energy losses   (\RAA and \RpPb) and the mechanism of  particle production. In these proceedings, we present recent results obtained in ALICE for the mesonic resonances (\rhmass, \kstarmass and \phimass) and baryonic resonances (\sigmass, \ximass and \lambmass) in \pp, \pPb, \PbPb and \XeXe collisions at various centre-of-mass colliding energies. Hereafter, the particles \rhmass, \kstarmass, \phimass, \sigmass and  \ximass will be represented by \rh, \kstar, \phim, \sigm and  \xim respectively.
\section{Analysis Method}
\label{sec-1}
This analysis used the data recorded in the ALICE detector~\cite{6}. The main components of the ALICE which are relevant to the analysis are Inner Tracking System (\ITS, for tracking and vertex finding), Time Projection Chamber (\TPC, for tracking and particle identification) and \VZERO detector (for triggering and centrality estimation). The events selected for the analysis are required to have a reconstructed primary vertex within 10 $cm$ of the nominal interaction point along the beam direction (at the centre of the ALICE barrel). The measurements of the resonance production are performed at mid-rapidity (\yrange{0.5} in \pp, \PbPb and \XeXe collisions and \yran{0}{0.5} in \pPb collisions) as a function of charge particle density which is also measured at mid-rapidity. The invariant mass is reconstructed from the measured momenta of identified particles. The combinatorial background is estimated using event-mixing technique. The mixed-event distribution is then normalized to the unlike-charge distribution in the mass range 1.65 < $M_{pK}$ < 1.75 \GeVmass for \lambmass and then subtracted from unlike-charge  distribution in each \pt bin. The residual background is removed by fitting the invariant mass distribution with a sum of Voigtian and a polynomial function. The raw yields are calculated after background subtraction and  are corrected with detector acceptance, tracking efficiency of daughter particles and branching ratio. Table~\ref{table:tabResonance} shows the lifetime values, decay modes used for reconstruction and the branching ratios of some of the measured  hadronic resonances. 
\begin{table}[H]
\centering
\caption{Lifetimes, reconstructed decay modes and branching ratios of measured resonance particles}
\label{table:tabResonance}       
\begin{tabular}{lllllll}
 & \rhmass & \kstarmass & \sigmass & \lambmass &\ximass & \phimass  \\\hline
\bf{Lifetime (fm/c)} & 1.3 & 4.2 & 5.5 & 12.6 & 21.7 & 46.4  \\
\bf{Decay mode} & \pip\pim & \kap\pim & \lmb\pipm & $p$\kam & \ximinus\pip & \kap\kam \\
\bf{B.R. (\%)} & 100 & 66.6 & 87.0 & 22.5 & 66.7 & 49.2 \\\hline
\end{tabular}
\end{table}
\begin{figure}[H]
	\centering
	\sidecaption	
		\includegraphics[width=0.5\textwidth]{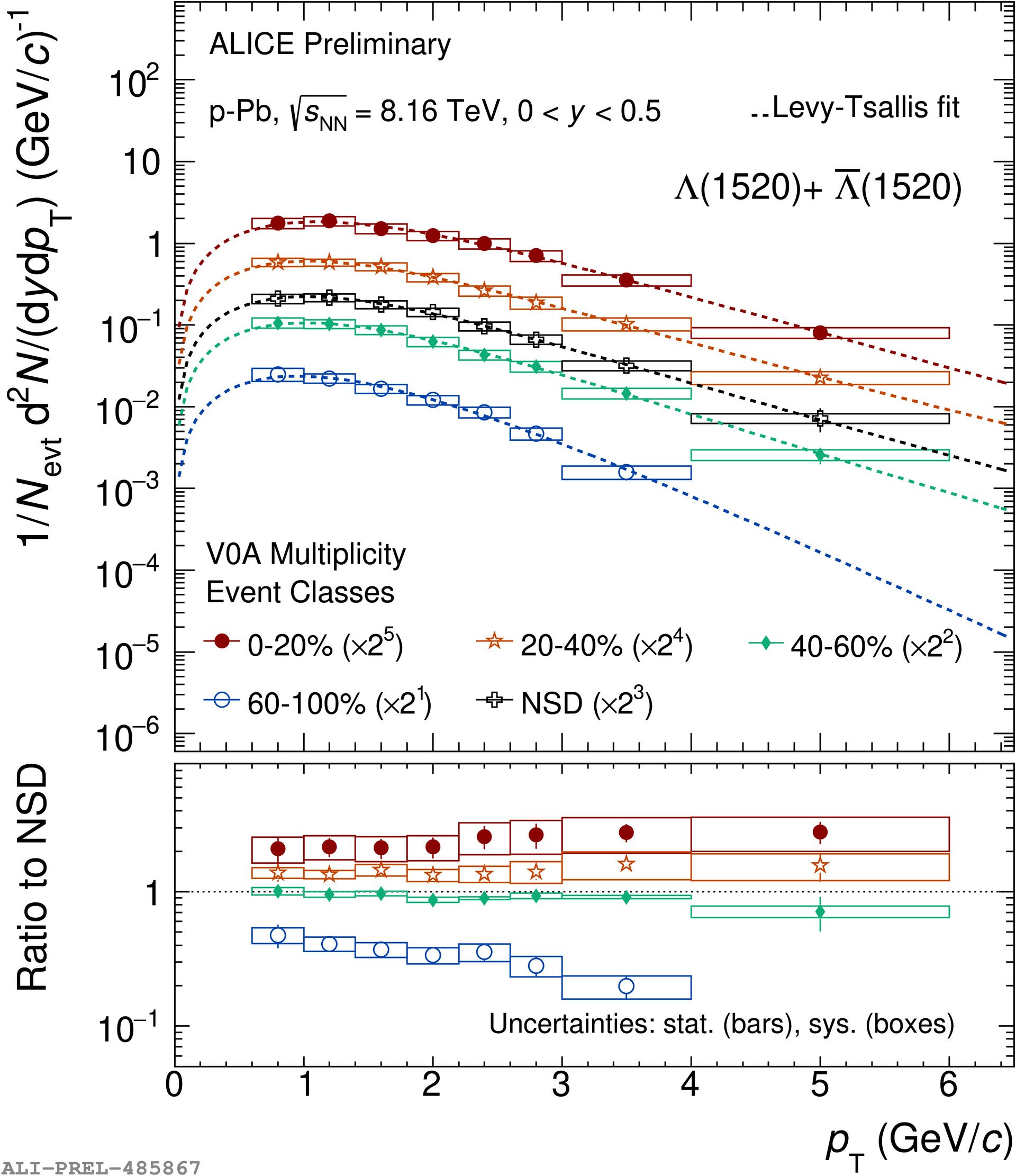}
		\caption{ Corrected \lambmass \pt spectrum for  minimum bias (0$-$100\%) and four multiplicity intervals (0$-$20\%, 20$-$40\%, 40$-$60\% and  60$-$100\%) in mid-rapidity p-Pb collisions at \snn=8.16 TeV. The errors shown here are both statistical (bars) and systematic (boxes). The shaded curves represent the Levy-Tsallis fit.}
		\label{fig:pTspectra}
\end{figure}
\section{Results and discussion	}
\label{sec-2} 
 We present new results on \pt spectra for \lambmass resonance in \pPb collisions in the mid-rapidity region at \snn = 8.16 \TeV in minimum bias (0-100 \%) and four \VZERO event multiplicity classes (0$-$20\%, 20$-$40\%, 40$-$60\% and  60$-$100\%) as shown in Fig.~\ref{fig:pTspectra}. We observed that \pt spectra harden with the increase in multiplicity and similar  behaviour is observed for other measured resonances  in \pp, \pPb, \PbPb and \XeXe collisions~\cite{7,8,9}.

\begin{figure}[H]
	\centering
	\includegraphics[width=0.45\textwidth]{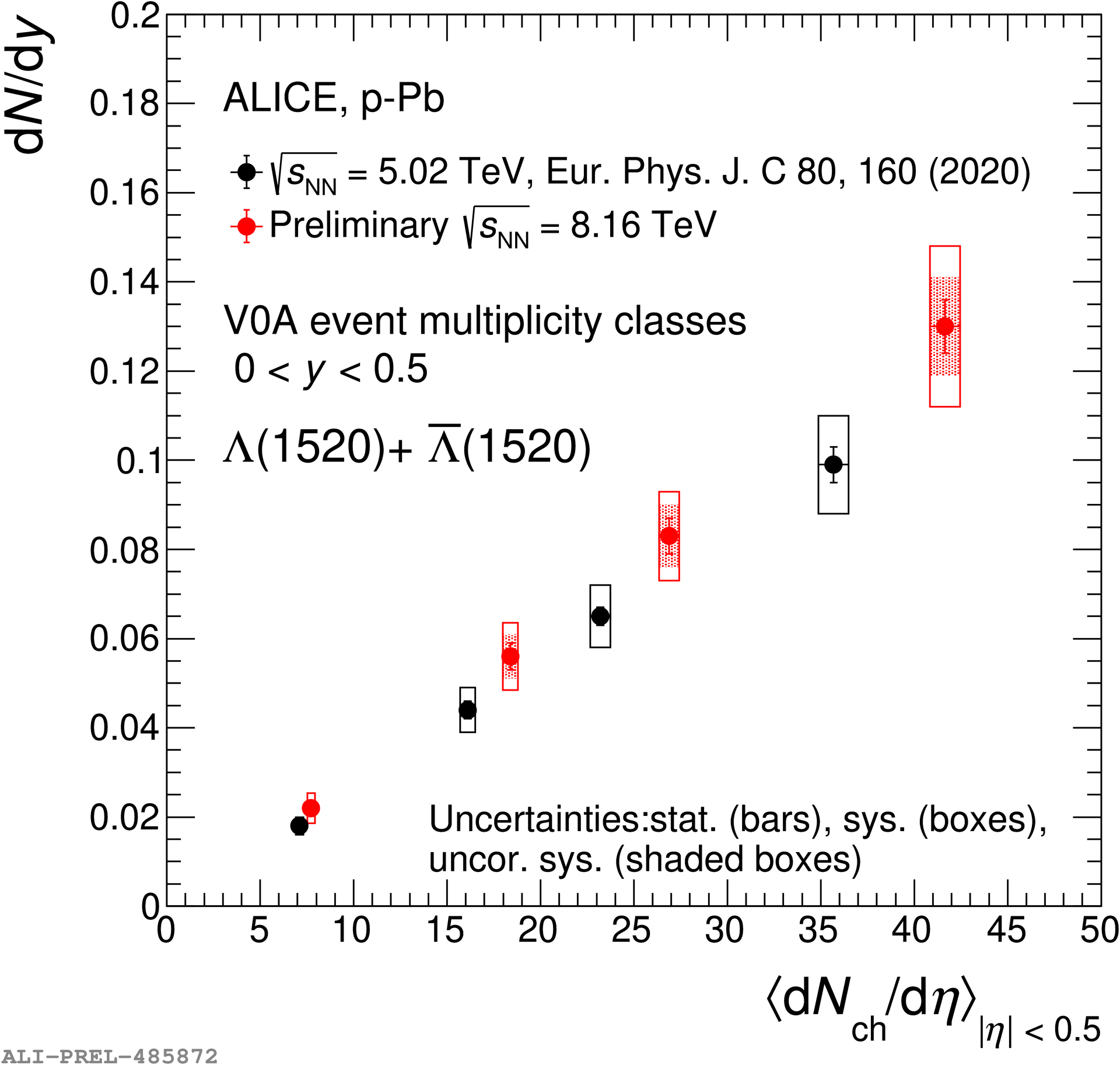}
	\includegraphics[width=0.45\textwidth]{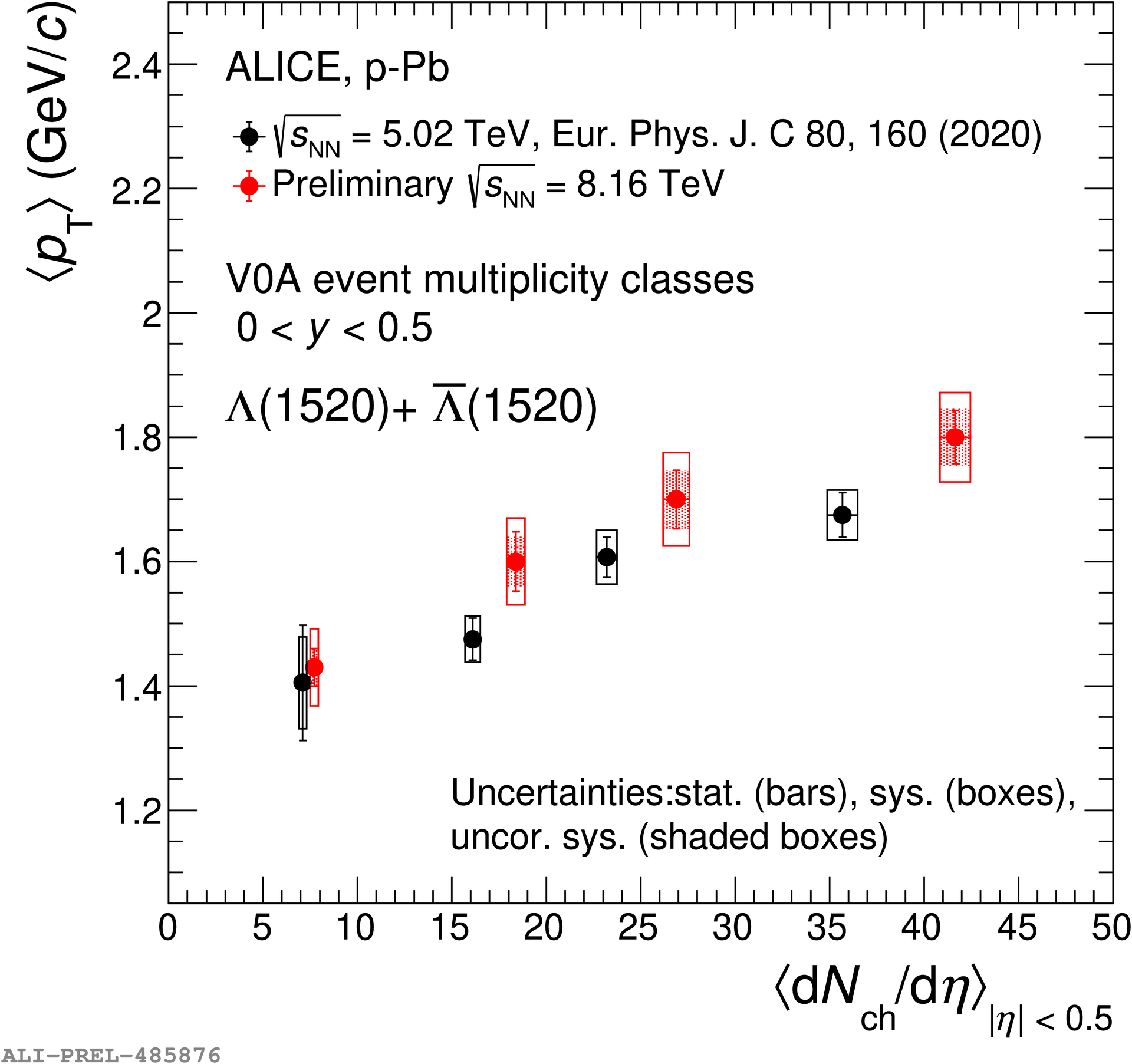}
	\caption{Integrated yield (left panel) and Average \pt (right panel) of \lambmass  in \pPb collisions at \snn= 5.02 and 8.16 \TeV  in mid-rapidity region in various \VZERO event multiplicity classes. Bars show statistical errors, boxes show systematic errors and shaded boxes show uncorrelated errors. }
	\label{fig:dNdy}
\end{figure}
The \pt integrated yields of the \lambmass in \pPb collisons at \snn= 5.02 and 8.16 \TeV (shown in the left panel of Fig.~\ref{fig:dNdy})  have been calculated in the mid-rapidity region in various \VZERO event multiplicity classes using the fitted Levy-Tsallis function to the measured \pt spectra of these resonances. Integrated yields of the resonances increase with event multiplicity and are independent of the colliding system and collision energy i.e. event multiplicity drives the resonance production.
For the first time, average \pt of \lstar in \pPb collisions at \snn= 8.16 TeV is measured in mid-rapidity region for various \VZERO event multiplicity classes (shown in the right panel of  Fig.~\ref{fig:dNdy}). The results have been compared with the previous measurements in \pPb at \snn= 5.02 \TeV~\cite{10} and are comparable within the systematic uncertainties. The average \pt increases with event multiplicity.

\begin{figure}[!tbp]
	\centering
		\includegraphics[width=0.8\textwidth]{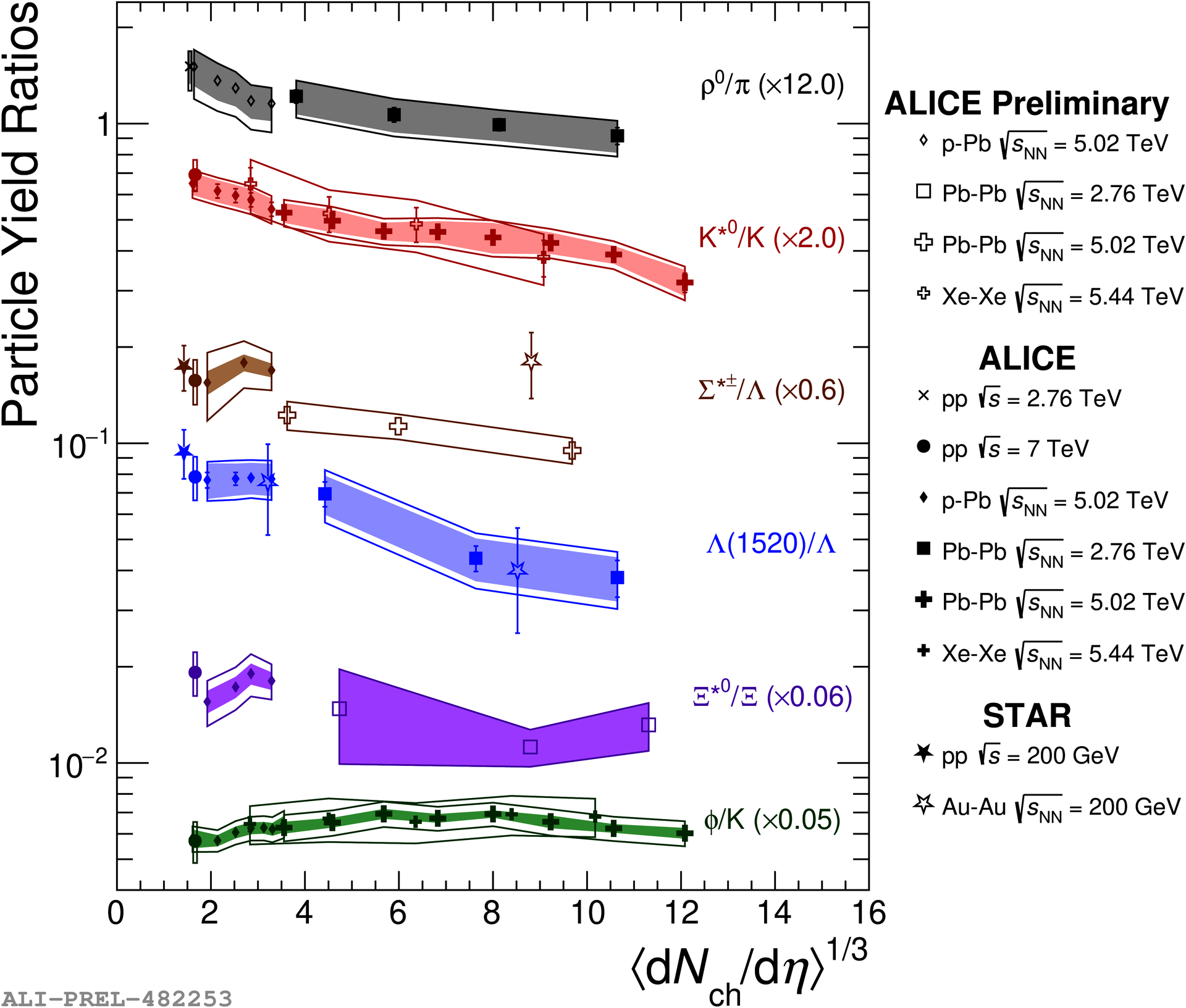}
		\caption{Ratios of integrated yields of various resonances to the stable particles in various systems and colliding energies.The bars and the boxes represent statistical and systematic uncertainty, respectively}
		\label{fig:Particleratios}
\end{figure}
The ratios of yields of resonances to those of stable hadrons is shown in the Fig.~\ref{fig:Particleratios}. We observe that the yields of \rh, \kstar, \sigm, \lambmass and \xim show a suppression with increase in the  centrality of \PbPb collisions while no such suppression is seen in the \phim resonance (due to its longer lifetime compared to the lifetime of hadronic phase). Different suppression of resonances, depending on their lifetimes, may provide some clue to estimate for a finite (non-zero as well as limited) lifetime of the hadronic phase. In small system collisions (\pp and \pPb), the yield suppression with increase in event multiplicity is only observed for \rh and \kstar resonances. This shows the dependence of the lifetime of the hadronic phase on the colliding systems i.e. the lifetime of the hadronic phase increases with the increase in the size of the colliding system.


\section{Conclusions}
\label{sec-3}
Recent results of the hadronic resonances (\rhmass, \kstarmass, \phimass, \sigmass, \lambmass and  \ximass) in \pp, \pPb, \PbPb and \XeXe collisions obtained from the ALICE detector have been presented.The \pt spectra of these resonances get harder with the increase in multiplicity of the events. At similar event multiplicities resonance yields are independent on the colliding system or collision energy, which indicates that multiplicity drives the production of resonance particles.  The lifetime of the hadronic phase is limited (non-zero and finite) and this lifetime increases with the increase in the size of the colliding system.

\end{document}